\documentclass{aa}
\usepackage{graphicx}
\begin{document}
   \title{Radiative transfer in cylindrical threads with incident
          radiation}

   \subtitle{VI. A hydrogen plus helium system}

   \author{P. Gouttebroze \inst{1} \and N. Labrosse \inst{2} }

   \institute{Institut d'Astrophysique Spatiale, Univ. Paris XI/CNRS,
              Bat. 121, F-91405 Orsay cedex, France\\
              \email{pierre.gouttebroze@ias.u-psud.fr}
        \and
              Department of Physics and Astronomy,
              University of Glasgow, Glasgow G12 8QQ, Scotland \\
              \email{n.labrosse@physics.gla.ac.uk}   }

   \date{Received 8 December 2008  / Accepted ................ }

\abstract
{Spectral lines of helium are commonly observed on the Sun. These
observations contain important informations about physical
conditions and He/H abundance variations within solar outer
structures.}
{The modeling of chromospheric and coronal loop-like structures
visible in hydrogen and helium lines requires the use of appropriate
diagnostic tools based on NLTE radiative tranfer in cylindrical
geometry.}
{We use iterative numerical methods to solve the equations of
NLTE radiative transfer and statistical equilibrium of atomic
level populations. These equations are solved alternatively for the
hydrogen and helium atoms, using cylindrical coordinates and
prescribed solar incident radiation. Electron density is determined
by the ionization equilibria of both atoms. Two-dimension effects
are included.}
{The mechanisms of formation of the principal helium lines are
analyzed and the sources of emission inside the cylinder are
located. The variations of spectral line intensities with temperature,
pressure, and helium abundance, are studied.}
{The simultaneous computation of hydrogen and helium lines, performed
by the new numerical code, allows the construction of loop models
including an extended range of temperatures.}

\authorrunning{Gouttebroze \& Labrosse}
\titlerunning{Radiative transfer in cylindrical threads. VI}
\maketitle
\section{Introduction}
Observation of the upper solar atmosphere with high angular resolution
reveals a wealth of filamentary structures produced by magnetic
fields. For modeling these objects, we developed a series of NLTE
radiative transfer codes that are described in the present series
of papers. Among the filamentary objects relevant to this kind of
modeling, we can mention:
\begin{itemize}
\item  cool coronal loops,
\item  chromospheric fine structure (cf. Patsourakos et al.
 \cite{PGV07}),
\item  prominence (or filament) threads (cf. Heinzel \cite{Hei07}),
\item  spicules.
\end{itemize}
Paper~I (Gouttebroze \cite{Gou04}) was dealing with 1D (i. e. radius
dependent) cylindrical models, a case applicable to cylindrical
structures with vertical axis, exposed to an incident radiation
field independent of azimuth. Papers~II and III (Gouttebroze
\cite{Gou05} and \cite{Gou06}, respectively) treated the case of
2D (radius and azimuth dependent) cylinders.
Paper~II was restricted to a 2-level atom, while Paper~III used a
multilevel hydrogen atom.
Papers~IV and V (Gouttebroze \cite{Gou07} and \cite{Gou08})
were dedicated to radiative equilibrium and velocity fields,
respectively. All these papers were dealing with the hydrogen atom.
Certain amount of helium was included in the state equation,
but it was assumed to be neutral and without any influence on the
radiation field, as well as on the electron density.\\
\\
In the present paper, we assume that the cylinders are filled with a
mixture of hydrogen and helium, and treat NLTE radiative transfer
and statistical equilibrium of level populations for both atoms
in two dimensions.
The helium model atom includes the three stages of ionization,
and the electron density is recomputed at each iteration in order to
satisfy the equation of electric neutrality.
In Sect.~2, we describe the computational methods used in the new
numerical code.
The results concerning hydrogen and helium ionization are detailed in
Sect.~3.
In Sect.~4, we study the formation of helium lines using a
reference model defined in Sect.~2.
Finally, in Sect.~5, we show how the helium line intensities react to
changes in temperature, pressure and helium abundance.
\section{Numerical methods}
  \subsection{Formulation}
The computation includes the numerical solution of the equations of
NLTE radiative transfer for hydrogen and helium atoms, statistical
equilibrium of level populations (for both atoms), pressure equilibrium,
and electric neutrality.
As in Papers~II and III, the object under consideration is a cylinder
of diameter $D$, whose axis makes an angle $\alpha$ with the local
vertical to the solar surface (this angle may vary between 0 and
90$\degr$).
The two active dimensions for radiative transfer are the distance to
axis $r$ and the azimuth $\psi$.
The method of resolution is of the MALI type (Rybicki and Hummer
\cite{RH91}).
The special form of equations for two-dimensional azimuth dependent
(2DAD) cylindrical geometry and the method of solution,
which are described in details in Paper~II, will not be repeated here.
The equations of statistical equilibrium, independent of geometry,
are treated in Paper~I.
In Paper~II, we also described the method to compute the intensities
incident on the cylinder, at different wavelengths, from the
knowledge of the emission by the Sun, the inclination $\alpha$, and
the altitude $H$. These incident intensities are also functions of
azimuth, except in the special case $\alpha = 0$.
The condition for electric neutrality may be written
\begin{equation}
N_{\rm e} = N_{\rm HII} + N_{\rm HeII} + 2~ N_{\rm HeIII}~,
\end{equation}
where $N_{\rm e}$ is the electron density, $N_{\rm HII}$ the number
density of ionized hydrogen (protons), and $N_{\rm HeII}$ and
$N_{\rm HeIII}$ the number densities of helium atoms in the second
and third stages of ionization, respectively.
If $N_{\rm H}$ and $N_{\rm He}$ are the total densities of hydrogen
and helium, respectively, and $N_{\rm HI}$ and $N_{\rm HeI}$ the
corresponding densities of neutral atoms, the law of particle
conservation yields
\begin{equation}
N_{\rm H} = N_{\rm HI} + N_{\rm HII}
\end{equation}
and
\begin{equation}
N_{\rm He} = N_{\rm HeI} + N_{\rm HeII} + N_{\rm HeIII}~.
\end{equation}
Besides, the gas pressure is
\begin{equation}
P_g = \left( N_{\rm H} + N_{\rm He} + N_{\rm e} \right)~k~T~,
\end{equation}
where $k$ is the Boltzmann constant and $T$ the temperature.
The parameter to check the convergence of iterations is the
electron-to-hydrogen ratio
$\omega = (N_{\rm e}/N_{\rm H})$.
If $A_{\rm He}$ is the (He/H) abundance ratio, Eq.~(4) becomes
\begin{equation}
P_g = N_{\rm H} \left( 1 + A_{\rm He} + \omega \right)~k~T~.
\end{equation}
The model atom for hydrogen includes 5 discrete levels plus 1 continuum.
The equations of radiative transfer for the 10 discrete transitions
and the 5 bound-free transitions are treated in details,
except in the case where the optical thicknesses are very low,
which generally happens for continua from subordinate levels.
The model atom for helium is the same as in Labrosse and Gouttebroze
(\cite{LG01}, \cite{LG04}). It contains 34 levels: 29 for He~I,
4 for He~II, and 1 for He~III.
The main part of this model comes from the
neutral helium model of Benjamin et al. (\cite{BSS99}). It is
complemented with parameters of various origins, as described in
Labrosse and Gouttebroze (\cite{LG01}).
The number of permitted radiative transitions is 76, but most of
them are optically thin under usual conditions.
All discrete transitions, for hydrogen as for helium, are treated
under the assumption of complete frequency redistribution.
\begin{table*}
\caption{Summary of cylindrical thread models}
\centering
\begin{tabular}{ccccccc}
\hline\hline
 name  &  $P_g$ (dyn cm$^{-2}$)  &  $T$ (K)  &  $A_{\rm He}$  &
$v_T$ (km s$^{-1}$)  &  $D$ (km)  &  nb. iter.  \\
\hline
t1  &  0.1  &  6000  &  0.1  &  5  &  1000  &  91$^*$ \\
t2  &  0.1  &  8000  &  0.1  &  5  &  1000  &  81   \\
t3  &  0.1  & 10000  &  0.1  &  5  &  1000  &  76   \\
t4  &  0.1  & 15000  &  0.1  &  5  &  1000  &  56$^*$ \\
t5  &  0.1  & 20000  &  0.1  &  5  &  1000  &  71$^*$ \\
t6  &  0.1  & 30000  &  0.1  &  5  &  1000  &  43   \\
t7  &  0.1  & 40000  &  0.1  &  5  &  1000  &   5   \\
t8  &  0.1  & 50000  &  0.1  &  5  &  1000  &   5   \\
t9  &  0.1  & 65000  &  0.1  &  5  &  1000  &   5   \\
t10  &  0.1  & 80000  &  0.1  &  5  &  1000  &   5   \\
t11  &  0.1  & 100000  &  0.1  &  5  &  1000  &   5   \\
p1  &  0.02 &  var.  &  0.1  &  5  &  2000  &  45   \\
p2  &  0.03 &  var.  &  0.1  &  5  &  2000  &  38   \\
p3  &  0.05 &  var.  &  0.1  &  5  &  2000  &  33   \\
p4 (or a4) & 0.1 & var. &  0.1  &  5  &  2000  &  42   \\
p5  &  0.2  &  var.  &  0.1  &  5  &  2000  &  40   \\
p6  &  0.3  &  var.  &  0.1  &  5  &  2000  &  39   \\
p7  &  0.5  &  var.  &  0.1  &  5  &  2000  &  51   \\
a1  &  0.1  &  var.  &  0.01 &  5  &  2000  &  41   \\
a2  &  0.1  &  var.  &  0.02 &  5  &  2000  &  41   \\
a3  &  0.1  &  var.  &  0.05 &  5  &  2000  &  42   \\
a5  &  0.1  &  var.  &  0.15 &  5  &  2000  &  42   \\
a6  &  0.1  &  var.  &  0.20 &  5  &  2000  &  43   \\
a7  &  0.1  &  var.  &  0.30 &  5  &  2000  &  43   \\
\hline
\end{tabular}
\end{table*}
  \subsection{Computational scheme}
The computation is organized along two parallel series of routines,
the one for hydrogen, the other for helium.
In each series of routines,
the variables dependent on the atomic structures as atomic parameters,
populations, intensities in different transitions, are gathered into
a specific COMMON, which is ignored by the main program. This main
program only contains geometrical ($D$, $r$, $\psi$, etc.) and
physical ($P_g$, $T$, $v_T$, etc.) variables and the populations
($N_{\rm HII}$, $N_{\rm HeII}$, etc.) necessary to determine the
ionization. The main phases of computations are:
\begin{itemize}
  \item
Initialization: determination of geometrical, physical and atomic
parameters. The incident intensities are also computed for each
position at the surface of the cylinder and each direction,
according to the method explained in Paper~II.
  \item
First evaluation of level populations, in the optically thin
approximation. By averaging the incident intensities, we obtain mean
intensities in the different transitions of hydrogen and helium.
From these intensities and physical parameters, we compute the
radiative and collisional transition rates. Then, we solve
statistical equilibrium equations to obtain atomic level populations
at each point of the ($r,\psi$) mesh. Since the transition rates depend
on electron density, it is necessary to iterate.
We start from an arbitrary value of $\omega$ (e. g. $\omega=0.5$) and,
using Eqs.~(3) and (5), successively deduce:
\begin{equation}
N_{\rm H} = \frac{P_g}{(1 + A_{\rm He} + \omega)~k~T}~,
\end{equation}
\begin{equation}
N_{\rm He} = N_{\rm H} A_{\rm He}~,
\end{equation}
and
\begin{equation}
N_{\rm e} = N_{\rm H}~\omega~.
\end{equation}
After computation of transition rates and solution of statistical
equilibrium equations, we compute $N_{\rm HII}$, $N_{\rm HeII}$, and
$N_{\rm HeIII}$ by adding the populations of individual levels
together, and deduce a new value of $\omega$ by
\begin{equation}
\omega = \frac {N_{\rm HII}+N_{\rm HeII}+2~ N_{\rm HeIII}} {N_{\rm H}}~.
\end{equation}
These operations are repeated until convergence.
  \item
Full iterations with radiative transfer.
This is the main part of the computation.
The external scheme is similar to that of the preceding step,
with a variable $\omega$ controlling
the convergence of iterations but, in the meantime, the internal
intensities for all transitions of hydrogen and helium are recomputed
according to the principles of NLTE radiative transfer: absorption
coefficients are derived from atomic level populations (determined
in the preceding iteration). Then, intensities are computed by
solving the transfer equation along each ray and integrating with
respect to direction and frequency. At the same time, the diagonal
terms of the $\Lambda$ operator are calculated by the method of
Rybicki and Hummer (\cite{RH91}).
The formulae appropriate to the cylindrical geometry are given in
Paper~II. The new intensities and the diagonal $\Lambda$ coefficients
are used to form preconditioned statistical equilibrium equations,
similar to those of Werner and Husfeld (\cite{WH85}).
These equations are solved to obtain new level populations.
Generally, one radiative transfer iteration for hydrogen and one
for helium, between two iterations on $\omega$, are sufficient to obtain
a good convergence. In a few cases, it was necessary to perform
two radiative transfer iterations for one $\omega$ iteration.
These cases are indicated by an asterisk in the last column of Table 1.
\end{itemize}
  \subsection{Models}
The model cylinders are defined by a series of geometrical and
physical parameters. The geometrical ones, diameter $D$, altitude $H$
and inclination $\alpha$, have been defined above. Physical
parameters include: gas pressure $P_g$, temperature $T$, microturbulent
velocity $v_T$, and relative helium abundance $A_{\rm He}$.
The code allows the definition of each of these parameters
as a function of $r$ and $\psi$, but this possibility is not used
in practice, except for the temperature.
A summary of models used in the present paper
is displayed in Table 1.
First, there is a series of isothermal models ("t1" to "t11")
with temperature ranging from 6000 to $10^5$ K.
Their diameter is fixed to 1000 km, their pressure to
0.1 dyn cm$^{-2}$, helium abundance to 0.1, and microturbulent
velocity to 5 km s$^{-1}$. Other models have a prescribed temperature
variation $T(r)$, which comes from Paper~III, and is represented in
the upper part of Fig.~1 (models using this temperature distribution
are indicated by "var." in the third column of Table~1). The mean
model "p4" (or "a4") is similar to that of Paper~III. In the series
"p1" to "p7", we change the pressure while keeping the other
parameters constant. In the series "a1" to "a7", we investigate the
effects of helium abundance. The last column of Table~1 indicates
the number of iterations necessary to achieve the convergence on
$\omega$, the criterion being fixed to $10^{-6}$ and the minimum
iteration number to 5.
\section{Ionization}
The introduction of helium ionization in models allows the electron
density to be greater than the hydrogen density.
Let $\xi=(N_{HII}/N_H)$ be the ionization ratio for hydrogen, and
similarly $\eta_1 = (N_{HeII}/N_{He})$ and
$\eta_2 = (N_{HeIII}/N_{He})$ be the corresponding ratios for the
two stages of helium ionization.
Equation~(9) becomes
\begin{equation}
\omega = \xi + A_{He}~\eta_1 + 2~A_{He}~\eta_2~.
\end{equation}
The ionization ratios $\xi$, $\eta_1$ and $\eta_2$ are principally
controlled by temperature.
However, near the surface of the structure, the influence of incident
radiation becomes more and more important and tends to moderate
the effects of temperature.
\subsection{Model with temperature gradient}
%
\begin{figure}
\centering
\includegraphics[width=\linewidth]{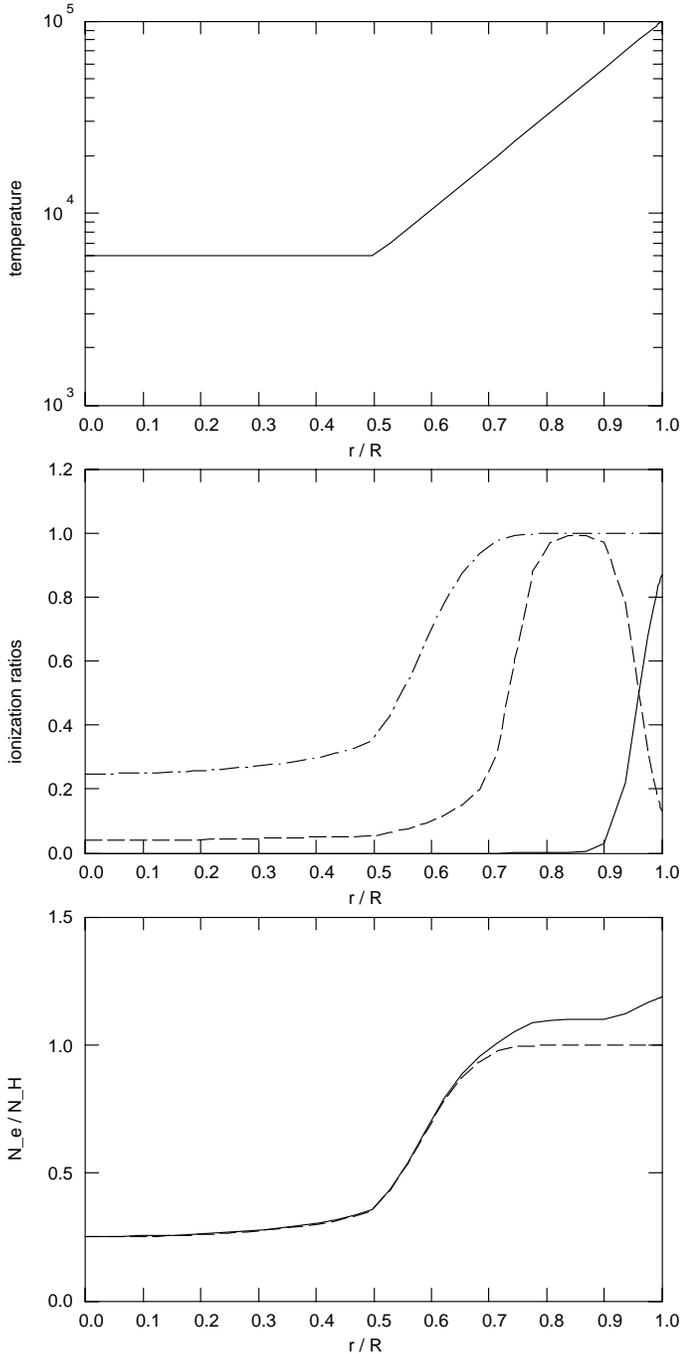}
\caption{ Variations of temperature $T$ and population ratios
with the distance to the axis ($r$),
for the model "p4" at the foot of the loop ($\alpha=0$).
Abscissae: distance to axis relative to the total radius $R$.
Top: temperature.
Middle: ionization ratios for hydrogen ($\xi$: dot-dashed line)
and helium ($\eta_1$: dashed line; $\eta_2$: continuous line).
Bottom: electron-to-hydrogen ratio $\omega$
(dashed line: model assuming neutral helium; continuous
line: model with both hydrogen and helium ionization). }
\end{figure}
%
In the standard model "p4", the temperature increases with the
distance to the axis, as shown in the upper part of Fig.~1.
In the core of the cylinder at 6000~K, helium is essentially neutral
and hydrogen is weakly ionized, with a ratio $\xi$ between 0.2 and 0.3.
The first ionization potential of hydrogen being lower than
that of helium, $\xi$ begins to increase first when moving towards
exterior.
This major change in hydrogen ionization occurs between
$0.5 R$ and $0.7 R$.
For $r > 0.7 R$, which corresponds to temperatures
$T > 20000$~K, hydrogen is almost completely ionized.
The first stage of ionization for helium (ratio $\eta_1$) occurs
between $(r/R)=0.7$ and $0.8$, i.e. temperatures between 20000 and
35000~K.
Finally, the second stage of helium ionization starts near $(r/R)=0.9$,
which corresponds to a temperature of 60000~K.
These variations of $\xi$, $\eta_1$ and $\eta_2$ are represented in
the middle part of Fig.~1.
The consequences for electron density are shown in the lower part of
Fig.~1, with a comparison to the case (Paper~III) where helium
ionization was neglected.
The two curves practically coincide up to $T=20000$~K, since hydrogen
is the only electron contributor there.
Between 30000 and 60000~K, $\omega$ remains close to 1.1,
the main part of helium atoms being singly ionized.
Near the surface, $\omega$ tends to 1.2, that is the possible
maximum value since we have assumed $A_{He} = 0.1$.
\subsection{Isothermal models}
%
\begin{figure}
\centering
\includegraphics[width=\linewidth]{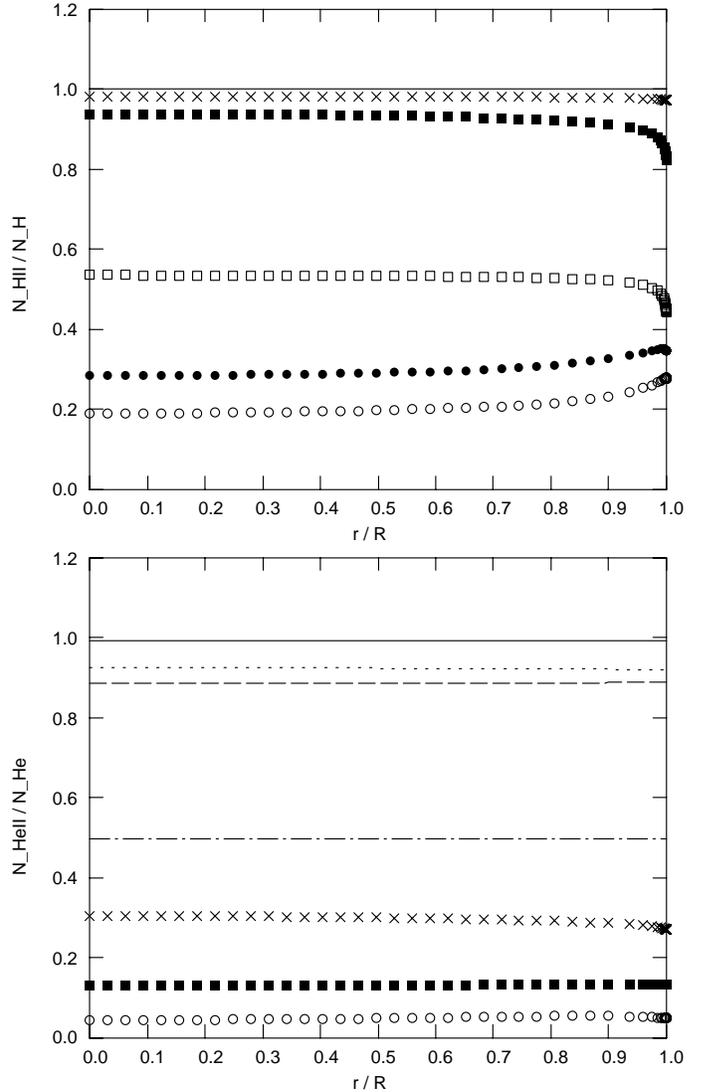}
\caption{ Variations of the ionization ratios $\xi$ (top)
and $\eta_1$ (bottom)
with $r$ for different isothermal models (for clarity, all models are
not represented).
Number densities are averaged with respect to the azimuth $\psi$.
Symbols for hydrogen:
open circles: $T = 6000$ K; full circles: 8000 K; open squares: 10000 K;
full squares: 15000 K; crosses: 20000 K; continuous line: 50000 K.
Same symbols for helium, plus: dotted line: 30000 K; dashed line:
65000 K; dot-dashed line: 80000 K. }
\end{figure}
%
In the standard model, the variations of temperature with $r$ masks
the effects of incident radiation on ionization.
To distinguish between thermal and radiative effects, it is necessary
to examine the variations of ionization within isothermal models.
We consider the series of "t" models described in Table~1.
Here, the cylinders have horizontal axes, but the number
densities are averaged over $\psi$ to obtain a single ionization curve
($\xi(r)$, $\eta_1(r)$ or $\eta_2(r)$) for each model and each ion
species.
Figure~2 represents the variations of ionization of hydrogen and
helium for the different models of this series.
It appears that the curves $\xi(r)$ and $\eta_1(r)$ are generally flat,
which confirms that the effects of temperature dominate those of
incident radiation.
This is especially true for helium.
For hydrogen, the $\xi(r)$ curves exhibit some inflexions near the
surface.
For the cool models at 6000 and 8000~K, the incident radiation tends
to increase hydrogen ionization near the surface.
In contrast, a decrease appears near the surface for models at
10000 and 15000~K.
For temperatures higher than 20000~K, the curves $\xi(r)$ are flat:
hydrogen is almost completely ionized, so that the cylinders are
optically thin in all transitions of the hydrogen atom, and
consequently insensitive to incident radiation.
Hydrogen is approximately half-ionized for $T=10000$~K.
The curves $\eta_1(r)$ corresponding to the first ionization stage
of helium are very flat, but a slight decrease near the edge may be
observed for the model at 20000~K.
The principal change of ionization for helium occurs between 20000~K
($\eta_1 \approx 0.3$) and 30000~K ($\eta_1 \approx 0.9$).
Between 30000 and 65000 K, more than 90\% of helium atoms are in
the first stage of ionization (nearly 100\% at 50000 K). At higher
temperatures, $\eta_1$ decreases while $\eta_2$ increases.
At 80000 K, the numbers of He\,II and He\,III ions are approximately
equal.
At 100000 K, $\eta_1$ is lower than 0.2, and consequently $\eta_2$
greater than 0.8.
The electron-to-hydrogen ratio $\omega$ (not represented) may be
easily deduced from the curves of Fig.~2, according to Eq.~(10).
At low temperatures, $\omega$ is nearly equal to $\xi$.
In contrast, above 50000 K, there are no longer neutral atoms,
so that $\xi = 1$ and $\eta_2 = 1 - \eta_1$, which gives
$\omega \approx 1 + 0.1 \times ( 2 - \eta_1 )$.
\section{Formation of helium lines}
The computations described in Sect.~2 provide us with absorption
coefficients and source functions in the different lines of hydrogen
and helium, which may be subsequently used to calculate the intensities
emerging from the cylinder.
The introduction of helium ionization has little influence on
hydrogen lines, which are formed in regions where helium is essentially
neutral.
Since hydrogen lines have been treated in preceding papers, we
concentrate here on helium lines, and use the standard model "p4"
to study their formation.\\
\\
\begin{figure*}
\centering
\includegraphics[width=\textwidth]{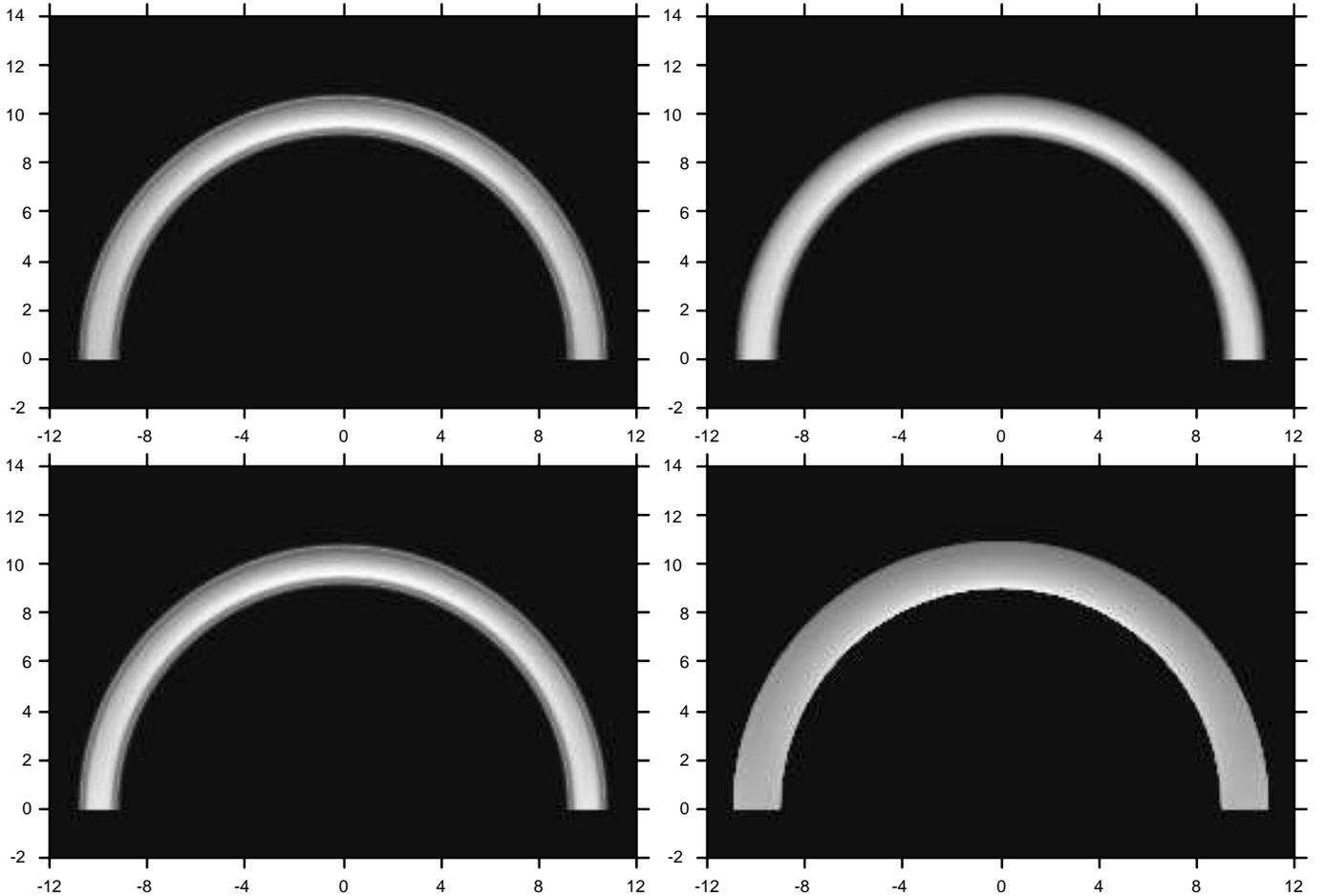}
\caption{ Emission of the loop model "p4" in several lines of
helium: He~I 10830~\AA\, (top, left); He~I 584~\AA\, (top, right);
He~I 5876~\AA\, (bottom, left); He~II 304~\AA\, (bottom, right).
Frequency-integrated intensities are normalized to the maximum value
of each image. Horizontal and vertical coordinates indicate
distances in megameters. }
\end{figure*}
The absorption coefficients and source functions for the different
values of $r$ and $\psi$, are computed for 7 values of the inclination,
from $\alpha=0$ to $\alpha=90\degr$.
Then, using an interpolation procedure described in Paper~III, we
construct a semi-toric loop model and obtain the intensities
emitted towards the observer.
Figure~3 shows the intensities, integrated over frequency, for
a few lines chosen among the most important of the helium spectrum.
The images corresponding to the two optical or infrared lines, say
10830 and 5876~\AA, look very similar to each other.
The consideration of absorption coefficients indicates that the
cylinder is optically thin for these two transitions (or marginally
thick for 10830 at line center).
In contrast, the two ultraviolet lines, He~I~584 and He~II 304~\AA,
are definitely optically thick. The structure seems broader in
He~II than in He~I lines, which is explained below.\\
\\
\begin{figure}
\centering
\includegraphics[width=\linewidth]{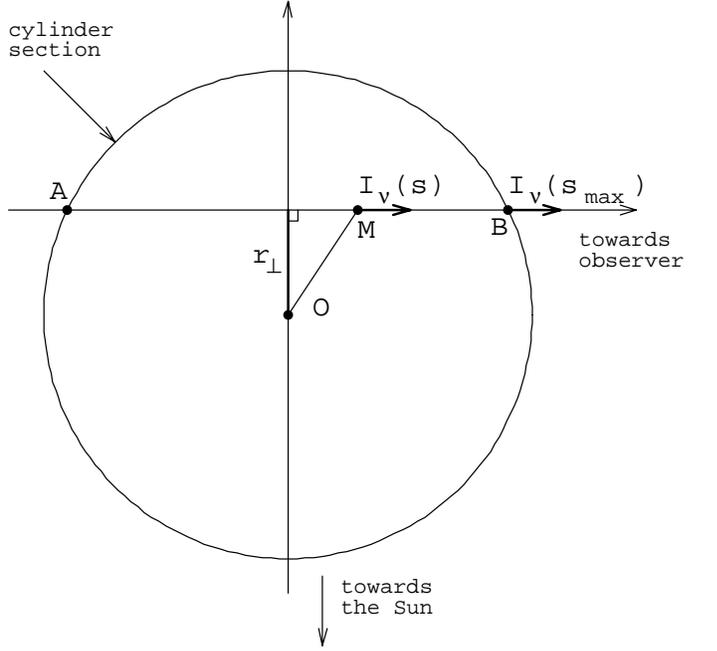}
\caption{ Ray path through the cylinder. }
\end{figure}
\begin{figure}
\centering
\includegraphics[width=\linewidth]{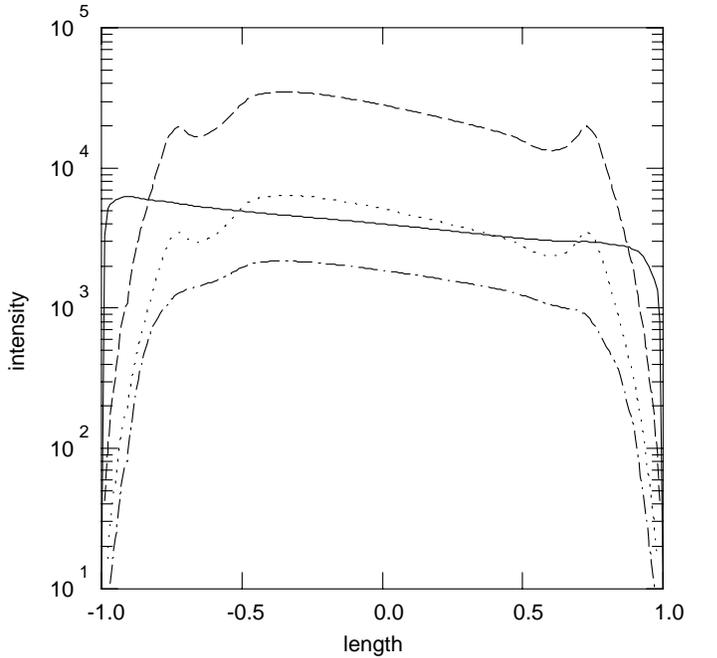}
\caption{ Total emergent intensities computed from model "p4",
for different helium lines vs. position across the cylinder.
Abscissae are in megameters and grow as the distance from the Sun.
Ordinates are intensities in erg cm$^{-2}$ s$^{-1}$ sr$^{-1}$.
Dashed line: He I 10830 \AA; dotted line: He I 5876 \AA;
dot-dashed line: He I 584 \AA; continuous line: He II 304 \AA. }
\end{figure}
To discuss the process of formation, it is sufficient to consider
the top of the loop, equivalent to a cylinder with horizontal axis.
For an observer looking at the cylinder in a direction normal to
the axis, the path of photons is represented on Fig.~4.
The ray enters the cylinder at point $A$ and exits at point $B$.
If $R$ is the radius of the cylinder and $r_{\perp}$ the distance
between the axis and the ray, the abscissae of $A$ and $B$ are
$-s_{\rm max}$ and $s_{\rm max}$, respectively, with
\begin{equation}
s_{\rm max} = \sqrt { R^2 - r_{\perp}^2 } .
\end{equation}
Let $\kappa_{\nu}$ and $S_{\nu}$ be the absorption coefficient and
the source function, respectively, at frequency $\nu$.
From the transfer equation
\begin{equation}
\frac { d I_{\nu} } { d s } = \kappa_{\nu} ( S_{\nu} - I_{\nu} )~,
\end{equation}
and the boundary condition
\begin{equation}
I_{\nu} ( - s_{\rm max} ) = 0~,
\end{equation}
we deduce the emergent intensity
\begin{equation}
I_{\nu} ( s_{\rm max} ) = \int_{-s_{\rm max}}^{s_{\rm max}}
 \kappa_{\nu}(s)~S_{\nu}(s)~e^{-\tau_{\nu}(s)}~ds~,
\end{equation}
with the optical thickness between the running point $M$ (abscissa $s$)
and $B$
\begin{equation}
\tau_{\nu} ( s ) = \int_{s}^{s_{\rm max}} \kappa_{\nu}(s')~ds'~.
\end{equation}
The total emergent intensity for the line under consideration is then
\begin{equation}
I_{\rm line} = \int_{\rm line} I_{\nu}(s_{\rm max})~d\nu~.
\end{equation}
These frequency-integrated intensities have been computed as functions
of $r$ and $\psi$ for different helium lines. They are shown in Fig.~5.
For all these lines, we observe a global decrease of intensities from
the lower to the upper edge, which is due to the decrease of incident
radiation. However, several differences may be noticed.
Concerning ultraviolet optically thick lines, the cylinder appears
broader in the He~II resonance line at 304~\AA\, than in the
corresponding line for He~I at 584~\AA.
Besides, the two triplet lines of He~I under consideration have
in common a transversal variation with three smooth peaks.\\
\\
\begin{figure}
\centering
\includegraphics[width=7cm]{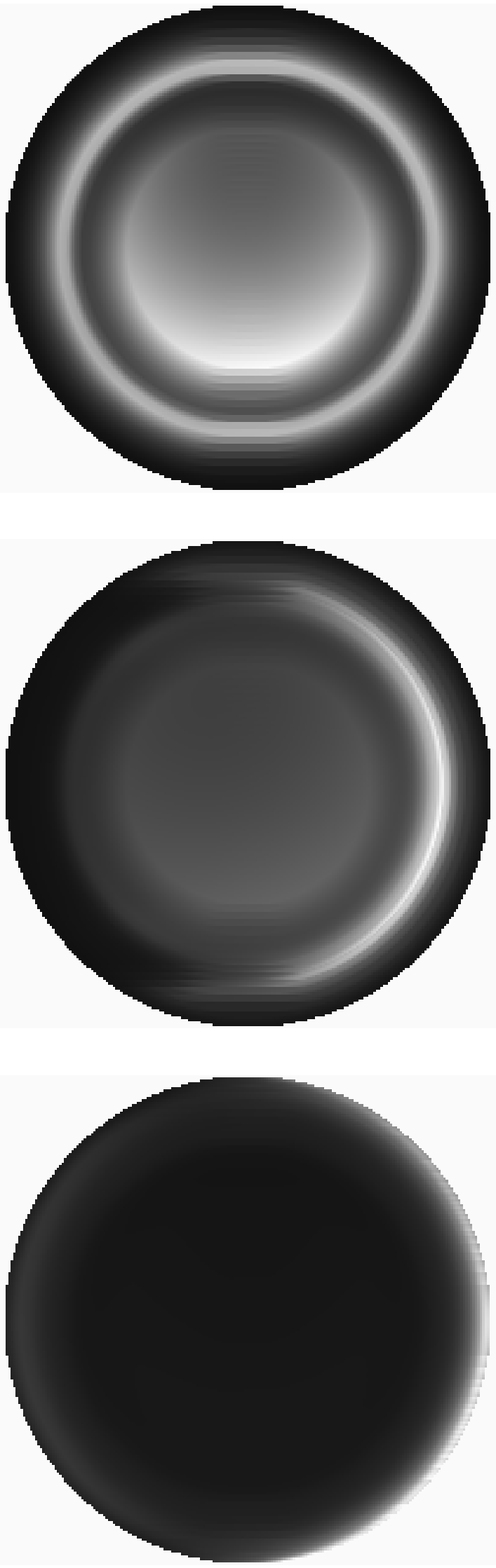}
\caption{ Contribution functions corresponding to the standard model
with horizontal axis, for three lines of helium.
Top: He I 10830 \AA; Middle: He I 584 \AA; Bottom: He II 304 \AA.
Bright zones correspond to the maximum of the function.
Directions are the same as in Fig.~4 (observer at right, Sun below). }
\end{figure}
%
To locate the origin of emissions inside the cylinder, we rewrite
Eq.~(14) as
\begin{equation}
I_{\nu} ( s_{\rm max} ) = \int_{-s_{\rm max}}^{s_{\rm max}}
 C_{\nu}(s)~ds
\end{equation}
with
\begin{equation}
C_{\nu}(s) = \kappa_{\nu}(s)~S_{\nu}(s)~e^{-\tau_{\nu}(s)}~.
\end{equation}
Thus, Eq.~(16) becomes
\begin{equation}
I_{\rm line} = \int_{-s_{\rm max}}^{s_{\rm max}} C(s)~ds
\end{equation}
with
\begin{equation}
C(s) = \int_{\rm line} C_{\nu}(s)~d\nu~.
\end{equation}
$C(s)$ is the contribution function appropriate for the
frequency-integrated emergent intensity.
This function is plotted as shades of gray in Fig.~6 for the three
lines at 10830, 584 and 304~\AA, respectively. The figure for the
5876~\AA\, line is very similar to that of the 10830~\AA\, line.
The existence of three maxima of emission for the intensity at
10830~\AA\, is due to the existence of two zones in the contribution
function: a central patch and a ring.
The patch corresponds to radiative processes of emission by
low temperature matter, while the ring corresponds to a range of
temperature where the maximum excitation of He~I is occuring.
The radiative processes of emission that occur in the central patch
include direct scattering of incident radiation and the
photoionization-recombination process (Hirayama \cite{Hir71},
Zirin \cite{Zir75}).
Since these two processes are dependent on incident radiation, the
contribution functions in the central zone decreases with height.
In contrast, the ring is produced by collisional excitation, so that
it is practically independent of $\psi$.
The contribution function for the 584~\AA\, line is concentrated in
a single ring, without central part: since the line is optically thick,
the front part of the ring only is contributing to intensity.
The same is true for the 304~\AA\, line but, in this case, the emitting
layer is located at high temperature, very close to the surface.
For this reason, the loop looks broader in this transition than in
the He~I lines.
\section{Influence of physical parameters}
\subsection{Temperature}
\begin{figure}
\centering
\includegraphics[width=\linewidth]{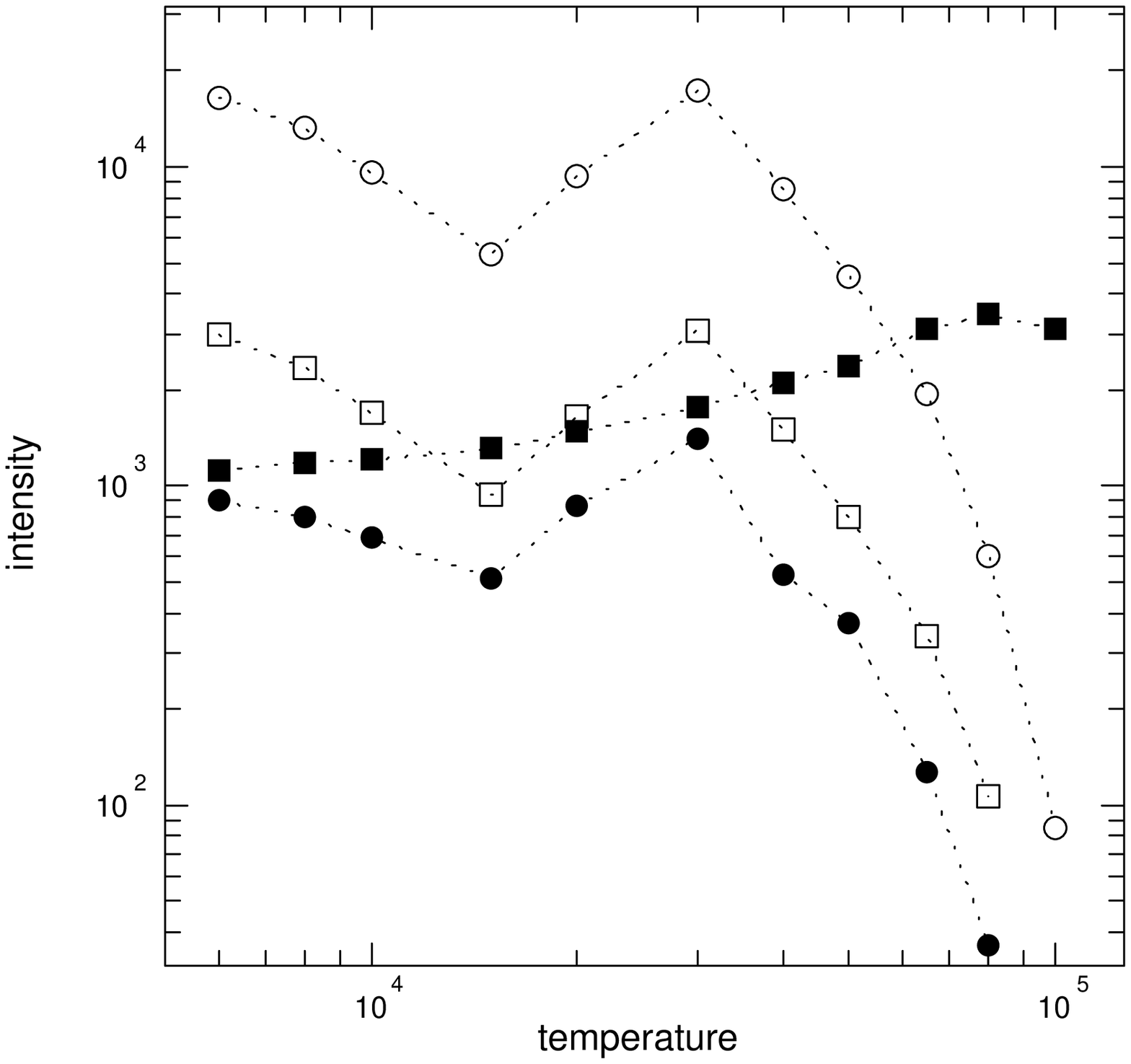}
\caption{ Frequency-integrated intensities, averaged over position,
emitted by isothermal models "t1" to "t11". Intensities
(erg cm$^{-2}$ s$^{-1}$ sr$^{-1}$) are plotted
as functions of the temperature (K) of the model, in 4 transitions:
He I 10830 \AA\, (open circles); He I 584 \AA\, (full circles);
He I 5876 \AA\, (open squares); He II 304 \AA\, (full squares). }
\end{figure}
%
To study the effects of temperature on emitted intensities, we use
the series of isothermal models ranging from "t1" (6000 K) to "t11"
($10^5$ K).
The cylinders under consideration have a horizontal axis
($\alpha=90\degr$) and the emitted intensities are averaged along
a transversal direction.
This spatial averaging is convenient for comparison with most
observations, where the width of the loops corresponds to a small
number of pixels.
Frequency-integrated intensities, for the different models and the
four principal lines, are displayed in Fig.~7.
The He~I triplet lines at 10830 and 5876~\AA\, have again a
similar behavior: the intensities decrease from 6000 to 15000~K, rise
from 15000 to 30000~K, and decrease again at higher temperatures.
As mentioned in the preceding section, these lines have two sources of
emission: radiative (dependent on incident radiation) and
collisional.
At low temperatures, scattering is dominant, and the decrease of
optical thickness results in a decrease of scattered radiation.
At higher temperatures, collisional excitation grows and, despite
the decrease of optical thickness, the emitted intensity rises.
For temperatures greater than 30000~K, the effect of opacity decrease
(principally due to ionization, as may be seen in Fig.~2)
dominates that of excitation increase, so that the intensity decreases
again.
The intensity variation of the resonance line at 584~\AA\, is similar,
but this line is optically thick at low or medium temperatures,
so that the slope of the curve between 6000 and 15000~K is not so
steep as that of the triplet lines.
At high temperatures, the decrease of the 584~\AA\, line intensity is
faster than that of the triplet lines, which may be due to the
decrease of density.
According to the analysis of neutral helium line formation by
Andretta and Jones (\cite{AJ97}), the intensity ratio (triplet/singlet
lines) decreases with increased density.
The behavior of the resonance line of He~II is quite different and
shows a slow and steady increase of intensity with temperature
until about 80000 K, followed by a decrease.
Most He~II lines reach their peak of emission between $5\times10^4$ K
and $10^5$ K (see for instance Laming and Feldman \cite{LF93} for
the 1640~\AA\, line).\\
\\
\begin{figure*}
\centering
\includegraphics[width=\textwidth]{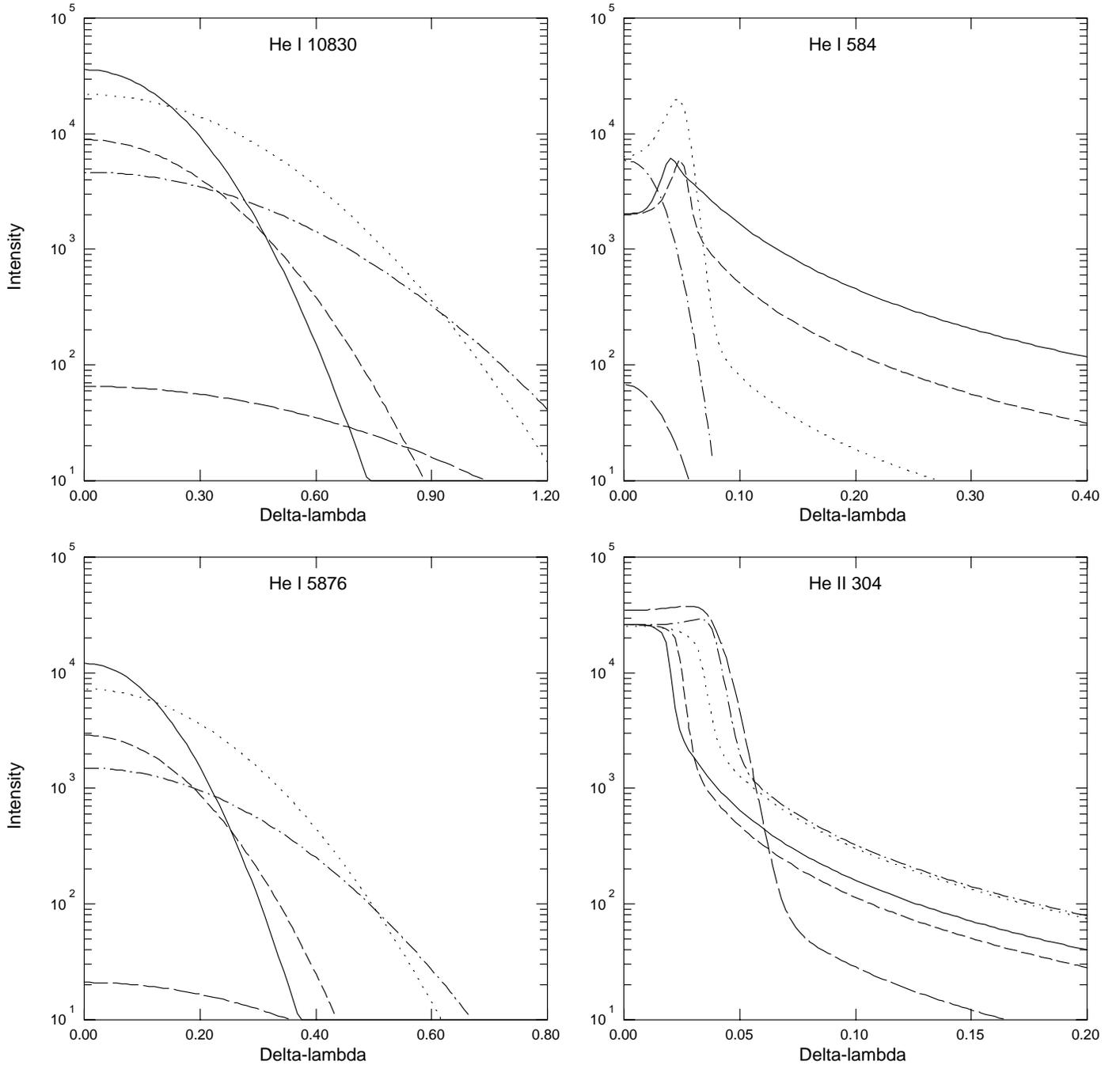}
\caption{ Helium line half-profiles emitted by five isothermal models:
continuous line: "t1" (6000~K);
dashed line: "t4" (15000~K);
dotted line: "t6" (30000~K);
dot-dashed line: "t8" (50000~K);
long-dashed line: "t11" (100000~K).
Each frame corresponds to a specific line, as indicated inside. }
\end{figure*}
%
The consideration of line profiles (Fig.~8) gives another insight on
the effects of temperature variations.
The profiles of triplet lines are nearly gaussian, a typical feature
of optically thin lines, their widths increasing with temperature.
The 584~\AA\, line exhibits a reversed profile from 6000 to
30000~K.
At 50000~K, the reversal disappears, indicating that the line is
optically thin.
This is due to the fast decrease of neutral helium populations
above $3\times10^4$ K (see Fig.~2).
At $10^5$ K, the central intensity of the 584 \AA\, line is two orders
of magnitude lower than at $5\times10^4$ K.
For the He~II line at 304~\AA, the line center intensity is roughly
constant until $5\times10^4$ K, and the global increase of intensity
is due to the progressive broadening of the line profile.
At these temperatures, the main process of emission is the scattering
of incident radiation, and the broadening of the line is due to the
joint effects of opacity (increase of $\eta_1$) and temperature
(Doppler effect).
Above $5\times10^4$ K, $\eta_1$ begins to decrease, but collisional
excitation grows.
These two competing effects produce a maximum of intensity near
$8\times10^4$ K.
In Fig.~8, one may compare the profiles of the 304 \AA\, line at
$5\times10^4$ and $10^5$ K: the higher temperature profile is more
intense at line center, but the intensity decreases more rapidly
in the wings, as a consequence of the smaller opacity.
\subsection{Pressure}
\begin{figure}
\centering
\includegraphics[width=\linewidth]{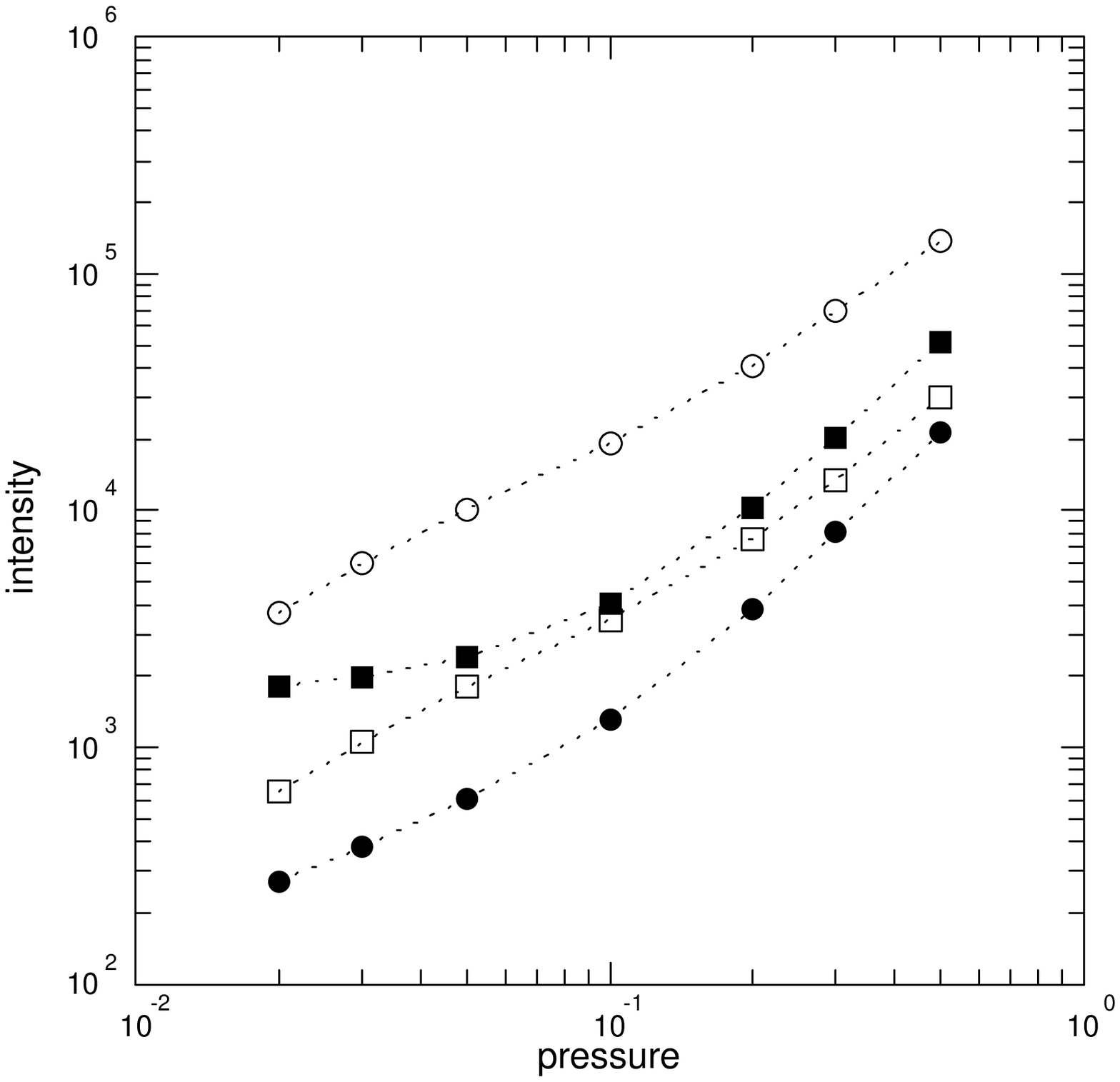}
\caption{ Frequency-integrated intensities, averaged over position,
emitted by models "p1" to "p7".
Abscissae: gas pressure inside the cylinder (dyn cm$^{-2}$).
Ordinates: intensities in erg cm$^{-2}$ s$^{-1}$ sr$^{-1}$.
Spectral lines:
He I 10830 \AA\, (open circles); He I 584 \AA\, (full circles);
He I 5876 \AA\, (open squares); He II 304 \AA\, (full squares). }
\end{figure}
The effects of gas pressure on emitted intensities are studied by
means of a series of models "p1" to "p7".
These models have the same variation of temperature as the standard
model, but differ from each other by pressures ranging from 0.02 to
0.5 dyn cm$^{-2}$.
Frequency-integrated intensities for the principal helium lines
emitted by these models are displayed in Fig.~9.
It appears that all intensities increase with pressure, but the slopes
of the curves differ from one line to another.
For neutral helium lines, the variation may be understood by
considering the two main processes of emission: scattering of
incident radiation and collisional excitation.
In the optically thin case, the first process yields intensities
proportional to atom populations.
The second emission process, collisional excitation, which is
proportional not only to the emitting atom density, but also to the
electron density, tends to produce a quadratic variation of emission
in function of pressure.
These considerations apply to the two He~I lines at 10830 and 5876~\AA.
It is visible in Fig.~9 that the emitted intensities for these two
lines are proportional to pressure from 0.02 to about 0.2 dyn cm$^{-2}$,
and that the slope slightly increases at higher pressures, when
collisions cease to be negligible.
The same considerations apply to the 584~\AA\, line, but the change of
slope begins at lower pressures, around 0.1 dyn cm$^{-2}$.
The case of the He~II 304~\AA\, line is more complicated, since the
slope of $I(P)$ is smaller than linear below 0.1 dyn cm$^{-2}$ and
greater than linear at higher pressures.
Besides, it is visible on profiles (Fig.~8) that this line is optically
thick and formed in the outermost part of the cylinder (Fig.~6).
Whatever the pressure, the incident ultraviolet radiation from the
Sun ionizes helium near the surface of the cylinder and creates a
zone which scatters the 304~\AA\, radiation.
This part of emission due to scattering is nearly constant and
constitutes the main contribution at low pressures (0.02 to
0.05 dyn cm$^{-2}$).
In contrast, at high pressures (0.2 to 0.5 dyn cm$^{-2}$),
collisional excitation becomes dominant and results in a
quasi-quadratic variation of $I(P)$.
\subsection{Helium abundance}
\begin{figure}
\centering
\includegraphics[width=\linewidth]{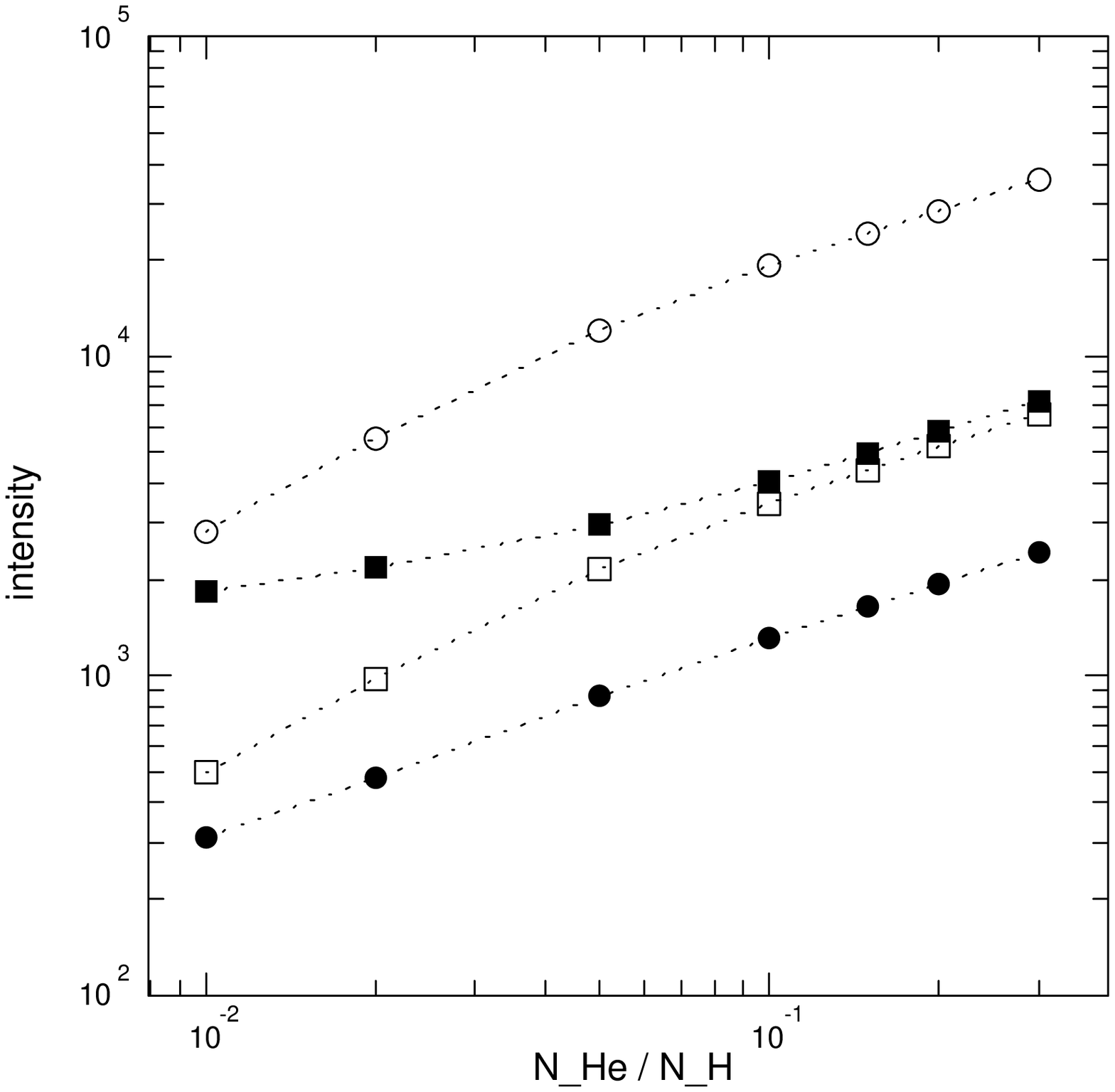}
\caption{ Frequency-integrated intensities, averaged over position,
emitted by models "a1" to "a7".
Abscissae: log (abundance ratio $A_{\rm He}$).
Ordinates: intensities in erg cm$^{-2}$ s$^{-1}$ sr$^{-1}$.
Spectral lines:
He I 10830 \AA\, (open circles); He I 584 \AA\, (full circles);
He I 5876 \AA\, (open squares); He II 304 \AA\, (full squares). }
\end{figure}
The abundance of helium is one of the most important parameters when
analysing helium lines (see for instance Andretta et al. \cite{AMF08}
and references therein).
Some observations, like that of filaments by Gilbert et
al. (\cite{GKA07}) suggest that very important changes of helium
abundance may occur between the top and the base of these objects.
To evaluate the importance of this parameter on the emission
of cylinder threads, we use a series of models ("a1" to "a7"), with
the same pressure and temperature variation as the standard model,
but helium to hydrogen ratios $A_{\rm He}$ varying from 0.01 to 0.3.
The variations of intensities as functions of $A_{\rm He}$ are
displayed in Fig.~10.
The interpretation of these curves is relatively easy.
As long as the cylinder is optically thin in the considered transition,
the intensity is proportional to $A_{\rm He}$.
This is the case for the 10830 and 5876 lines until $A_{\rm He}=0.05$.
For higher abundances, the slopes of the curves slightly decrease as
the line centers begin to saturate.
For optically thick lines at 584 and 304~\AA, the intensity is still
a growing function of abundance, but the slope is significantly lower
than that corresponding to proportionality.
\section{Conclusion}
At this stage of our project, it is possible to perform a modeling
of solar coronal loops including the following ingredients:
\begin{itemize}
\item NLTE radiative transfer in cylindrical geometry,
\item 2 dimensions (radius and azimuth),
\item statistical equilibrium of atomic level populations,
\item self-consistent treatment of the two principal chemical elements:
hydrogen and helium, including ionization,
\item pressure equilibrium.
\end{itemize}
The numerical code produces in a single run spectral line profiles
and intensities for the principal transitions of hydrogen and helium,
emitted in all directions.
Some developments remain to be done, as the implementation of motions
in helium routines (as was done for hydrogen in Paper~V), or the
treatment of Lyman lines in partial redistribution.
We also are willing to apply our code to the modeling of complex
loop systems observed in the solar atmosphere.
In this scope, we envisage to extend the code to some minor chemical
species, in function of the contents of observations.
\begin{acknowledgements}
We wish to thank Petr Heinzel for useful suggestions concerning this
manuscript.
\end{acknowledgements}

\end{document}